\documentstyle[11pt,newpasp,twoside,epsf]{article}
\def\edcomment#1{\iffalse\marginpar{\raggedright\sl#1\/}\else\relax\fi}
\reversemarginpar

\begin{document}
\title{Simulations of Galaxy Clusters observed by Chandra
}
\author{Alessandro Gardini}
\affil{Dipartimento di Astronomia, Universit\`a di Padova,
vicolo dell'Osservatorio 2, 35122 Padova - ITALY}

\begin{abstract}
A software package able to simulate imaging observations of galaxy 
clusters by the Chandra X-ray telescope is here presented.
We start from high resolution N-body hydrodynamical 
simulations of galaxy clusters and assign to each gas particle 
a spectrum of emissivity, after assuming the MeKaL model. 
We then construct spatial images of the source differential flux 
which are used to create lists of incoming X-ray photons, preserving
information on photon direction and energy.
The photon lists are passed on to the Chandra simulator (MARX) 
to produce the final observation events. 
Background events are added to complete the simulation. 
Data analysis is currently in progress and simulated observations 
by other telescopes will become available in the Future.
\end{abstract}

\section{Introduction}

We present a software package that, starting from high resolution 
hydrodynamical simulations of galaxy clusters, reproduces observations 
of the ICM by the Chandra X--ray telescope.
Syntetic data can then be processed and analysed using the same tools
and procedures as real observations.
This enables us to directly compare the results of the simulations with the 
observational data and furthermore to compare the intrinsic properties of 
clusters, as given by simulations, with the same quantities obtained at the
end of the process of data reduction and analysis.
In this way, we are able to check the reconstructed properties of
clusters as well as to get indication on the quantity of information 
retained or lost in the observations. 
On the other hand it has to be remembered that the reliability of
the comparison between a simulated quantity and the real counterpart
depends on how well the relevant physics inside simulations is modelled.

\section{Cluster Simulations}

The package can be easily adapted to run on every kind on hydro-simulations.
For our input we used a set of cluster simulations
realized by G.Tormen via the resimulation technique of regions of interest
inside a cosmological simulation 
(see e.g. Tormen, Bouchet, \& White 1996).

This starting simulation reproduces a $\Lambda$CDM model of Universe, with
parameters $\Omega_0=0.3$, $\Omega_\Lambda=0.7$, $H=70\,km\,s^{-1}\,Mpc^{-1}$
and baryon fraction $\sim 0.1$. It describes a box of size equal to
479 $Mpc\,h^{-1}$ with $512^3$ particles of Dark Matter\footnote{
http://www.mpa-garching.mpg.de/Virgo/VLS.html}.

Clusters are identified in its final output using a spherical overdensity
algorithm and resimulated singularly from the beginning using the GADGET 
Tree-SPH code. Each resimulation makes use of $\sim 10^6+10^6$ particles of 
Dark Matter and gas to model the cluster and the surrounding region, while
the external tidal field is reproduced by a smaller number of more massive
particles. The mass of the Dark Matter particles ranges between 2 and 
$5\cdot 10^9 M_\odot \,h^{-1}$ depending on the cluster, and the gas particle
mass scales accordingly. 
The softening related to the cubic spline kernel is $\sim 5\, kpc\,h^{-1}$. 
At the end of the simulation the cluster is described by 2 to $3 \cdot 10^5$
particles for each species inside the virial radius.
A set of twelve clusters has been simulated and for each one 50 outputs 
were retained between $z=10$ and $z=0$.

\begin{figure}
\vbox to8.cm{\rule{0pt}{0.cm}}
\includegraphics{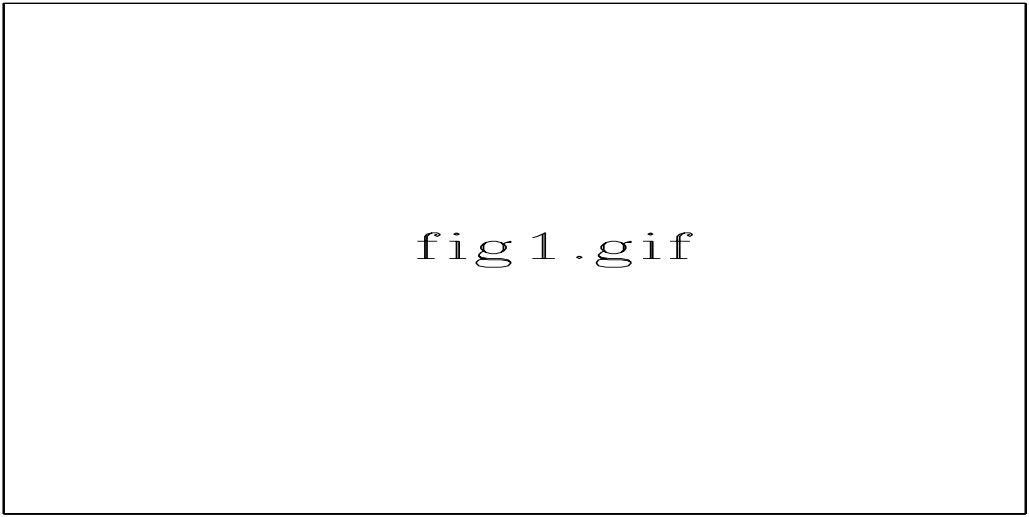}
\caption{
Left: map of projected flux ($ photon \,s^{-1}\,cm^{-2}$) per pixel
of a cluster at redshift $z\simeq 0.778$. 
Right: simulated observation of the same object by the Chandra 
ACIS-S3 detector given an exposure time of 19500 $s$. 
The image is turned upside-down and the main clump is set at the 
Chandra ACIS-S aimpoint.
}
\end{figure}

\section{Method : (1) general}

In order to describe how the package works, it can be
useful to distinguish between a first general part of the method
and a second one related to Chandra.
For a cluster at a given redshift $z$, the first part accounts for 
the construction of a map of the incoming flux (or of some other 
related quantity) that preserves the energy information.

We start by considering the position $\vec x_i$, mass $m_i$, density $\rho_i$ 
and temperature $T_i$ for each gas particle $i$. We then assign to the
particle a volume $V_i=m_i/\rho_i$ that is assumed cubic and centered on
$\vec x_i$. 

We then assign to each particle the related spectral emissivity 
$\epsilon^{\nu}_i=\epsilon^{\nu}_i(\rho_i,T_i)$ computed by XSPEC
using the MeKaL spectral model. The metallicity enters as a 
parameter; here we present results obtained assuming $Z=0.3\,Z_\odot$,
but any other value is possible. The luminosity of each particle becomes 
then $L_i^{\nu}=\epsilon^{\nu}_i\cdot V_i$
and is assumed uniformly distibuted on $V_i$.

In general, we consider emissivity and derived quantities 
(luminosity, flux, surface brightness) in terms of incoming photons 
instead of energy because this quantity is more easily related to observations. 

Given an observational direction, we get a bidimensional map of the
cluster spectral luminosity projecting each $L^{\nu}_i$ on a pixellized
perpendicular image and summing on particles. The information is then
stored in a 3D array, where two dimensions are spatial and the third one
contains the energy channels. The transformation from the cluster luminosity
at redshift $z$ to the flux per pixel at $z=0$ is then direct:
in term of incoming photons, it holds
$$ F_{\nu} = {(1+z)^2 L_{\nu (1+z)} \over 4\pi D_L^2},
$$
being $D_L$ the luminosity distance. Note that the frequency ${\nu}$
in the flux is related to $\nu (1+z)$ in the cluster luminosity.

Typically, the number of pixels ranges from $256^2$ to $1024^2$,
depending on the field of view, 
while the energy ranges from $0.1$ to $10 \, keV$ in bins of 
$20 \, eV$, even if $100 \,eV$ should be sufficient 
according to the energy resolution of Chandra (Chandra Proposers'
Observatory Guide\footnote{http://asc.harvard.edu/udocs/docs/docs.html}).

\section{Method : (2) Chandra}

It has to be remembered that a simulator of Chandra already exists: 
it is called MARX\footnote{http://space.mit.edu/ASC/MARX} 
and performs the ray-tracing of incoming
photons of a given source inside the telescope up to the possible 
final detection. It returns a list of binary data files describing 
the detection events that can be assembled to reproduce a CXC Level 1
FITS data file.
 
However, only some kind of sources of simple form and uniform spectral 
flux are implemented in MARX, while users are allowed to construct their 
specific source writing some C routines and linking it dynamically
to the code. So did we, as described in the following.

For a given cluster, 
we generate a list of incoming photons by random extraction from the
cumulative distribution of the spectral flux array described above.
Each photon hence owns a position in a pixel and an energy in the energy band, 
with the given resolution.
Other random extractions define the position of photon inside the pixel
and its energy in the bin. The number of photons extracted is set to $10^6$.

We then enable MARX to accept as input this photon list and the total 
flux of the cluster (in $photon\,s^{-1}\,cm^{-2}$): photons are randomly
extracted from the list according to the total flux and the exposure time,
and usually processed up to the possible detection. 

No background is introduced by MARX. We add background events randomly
extracting them from Chandra ACIS observations of empty fields available
in the home page of 
M.Markevitch\footnote{http://hea-www.harvard.edu/$\sim$maxim}.
An example of a complete simulated observation is shown in the Fig. 1.

\section{Data analysis and future developments}

As written above, the detection events generated by MARX have the same
caracteristics of the CXC Level 1 observational data file and can be
analysed using the same procedures and CIAO tools.
This analysis is currently in progress and will concern the reconstucted
properties of clusters, the cluster morphology (maps and profiles)
and its relation to the dynamical state of the object. 
Detectability of clusters at high redshift is another possible application.

We plan to extend this package to the XMM-Newton telescope, using the related 
simulator SCISIM, in the next future.

\section{Acknowledgments}
The author wish to thank L.Moscardini, G.Tormen and S.DeGrandi, 
that collaborate to this project.


\begin{references}
\reference Tormen, G., Bouchet, F.R., \& White, S.D.M., 1997,
\mnras , 286, 865
\end{references}
\end{document}